\documentclass{article}
\usepackage{amssymb}
\usepackage{amsfonts}
\usepackage{graphicx}
\usepackage{amsmath}

\newtheorem{theorem}{Theorem}[section]

\newtheorem{conjecture}[theorem]{Conjecture}

\begin{document}

\title{The importance of probability interference in social science:
rationale and experiment}
\author{Andrei Yu. Khrennikov\thanks{%
Director - International Center for Mathematical Modelling in Physics and
Cognitive Sciences - MSI- University of V\"{a}xj\"{o} - Sweden} and Emmanuel
Haven\thanks{%
School of Management - University of Leicester - United Kingdom}}
\maketitle

\begin{abstract}
Probability interference is a fundamental characteristic of quantum
mechanics. In \ this paper we attempt to show with the help of some
examples, where this fundamental trait of quantum physics can be found back
in a social science environment. In order to support our thesis that
interference can possibly be found back in many other macro-scopic areas, we
proceed in setting up an experimental test.
\end{abstract}


\section{Introduction}

The concept of `probability interference' is one of the most fundamental
concepts of quantum mechanics. Quantum mechanics, as a discipline of
physics, models quantum mechanical phenomena which typically are found at
scales well below the nuclear level. Hence, it may come as a surprise that
quantum physical methodology can be used in a setting which is substantially
far removed from the quantum mechanical scale: i.e. the macro-scopic scale.
In this paper we attempt to convince the reader that the use of precisely
such methodology can be used in a beneficial way in a social science
setting. More precisely, in this paper, we focus our efforts to purposefully
show how basic quantum physical concepts can be used to explain
experimentally observed violations of a very basic principle in economic
theory: the so called sure-thing principle\footnote{%
This principle is a key building block in some expected utility theories.
Please see section 3.}.

In order to begin to achieve this purpose we must first explain what
probability interference is. Then, after we will have explained the
essential meaning of the sure-thing principle, we proceed to illustrating,
following the path-breaking arguments by Busemeyer et al. (2006 and 2007),
how the violation of this principle can be explained with the help of
probability interference. In this paper we also highlight two important
paradoxes, the so called Allais and Ellsberg paradoxes, where the latter
paradox is an example of where the sure thing principle goes wrong. We
briefly illustrate how the Allais paradox is related to the `double slit'
experiment, which is the fundamental experiment used in quantum mechanics to
argue for the presence of probability interference.

To further underline the importance of the Ellsberg paradox and its
connections with quantum mechanical tools, we set ourselves two further main
aims in this paper.

As a first further main aim, we proceed first to show how information can be
modelled via the wave function - a central concept in quantum mechanics. We
then attempt to show that ambiguity, a main characteristic of the Ellsberg
paradox, can be modelled with the help of such wave function. Following
this, we also highlight how `ignorance' about ambiguity can be related to
the same basic quantum mechanical concept.

As a second further main aim, we would like to propose how we can test for
the existence of probability interference in a macro-scopic setting. If the
proposed test can be seen as a potentially conclusive test by which such
interference can be experimentally verified, we will have added one more
argument in support of why the violation of the sure-thing principle could
indeed be explained by probability interference.

In the next section of the paper we acquaint the reader with the notion of
the double slit experiment and the ensuing probability interference. In
section 3 of the paper, we first discuss the sure-thing principle and the
related Ellsberg paradox. We also explain the Allais paradox and briefly
argue how this paradox is related to the double slit experiment (the
experiment which triggers the existence of probability interference). In
section 4, we highlight, following the important work of Busemeyer et al.
(2006 and 2007), how probability interference can help in explaining the
violation of the sure-thing principle. In section 5, we define `ambiguity'
via the Ellsberg paradox. In section 6, we attempt to show how information
can be modelled via the use of the wave function. In section 7, we then
proceed in showing how ambiguity can be modelled via the wave function. In
section 8, we show how ignorance about ambiguity can again be modelled by
this same device. In section 9, we provide for a detailed description of an
experimental test we propose to prove the existence of probability
interference.

\section{The double slit experiment and probability interference}

The notion of probability interference is closely tied to one of the most
fundamental experiments in quantum physics: the double slit experiment. This
experiment consists of an electron gun which basically produces a beam of
electrons, and an electron detector. The detector counts the number of
electrons landing on a given area. We imagine there are two slits of equal
width. Both slits, let us call them slits $A$ and $B$, are separated by a
certain distance from each other. The experiment consists of three scenarios:

\begin{enumerate}
\item slit $A$ is open and slit $B$ is closed

\item slit $A$ is closed and slit $B$ is open

\item slit $A$ is open and slit $B$ is open
\end{enumerate}

Before any electron gun is activated, let us first imagine what would happen
if the electron gun were instead to be a gun firing out very small plastic
balls. We imagine those balls diameter to be very small compared to the
length of each slit. In scenario 1, when the plastic ball gun is activated,
we would expect that the plastic balls would pile up behind slit $A$ and
very few plastic balls would pile up behind slit $B$ (some would pile up
there because they would be scattered at the edges of slit $A$). The
opposite pattern would occur for scenario 2. When both slits are open we
would imagine there would be a density of balls piling up behind slit $A$
and a comparable density of balls piling up behind slit $B$. Some density of
balls would occur in between the slits because some of the balls would be
scattered at the edges of each of the slits $A$ and $B$.

The big question, loosely speaking then, becomes whether if with the
electron gun one will observe the same pattern? We follow here Morrison
(1990). Assume that both slits $A$ and $B$ are open. It occurs that once the
electron gun is opened, initially spots form, randomly behind the slits.
However, after some time the electrons start forming an interference
pattern. As is remarked in Morrison (1990) at first, the electrons behave
like particles. However, when time moves on they start behaving like waves.
Even if the electron gun fires off one electron at a time (passing through
one of the slits) there is still interference (i.e. the electron interferes
with itself)! This experiment, with the electron gun, shows that electrons
go through both slits. Interference forms even when one electron goes
through a slit. This result is indeed resolutely remote from what we would
expect if the electrons were plastic balls.

Quantum physicists devised a probability formula to describe those quite
mysterious events. We still follow Morrison (1990)\footnote{%
We slighlty alter the notation from Morrison (1990).}. Busemeyer et al
(2006) also provide for a very intuitive background on the discussion which
is now following.

Let us denote with $p_{A}(x,t)$ the probability that the electron arrives at
position $x$ at time $t$ when slit $A$ is open and similarly for $p_{B}(x,t)$%
. What would be the expression for the probability of finding an electron at
position $x$ at time $t$ when both slits are open? I.e. what is $%
p_{AB}(x,t)? $ Will it be:

\begin{equation}
p_{AB}(x,t)=p_{A}(x,t)+p_{B}(x,t)?
\end{equation}%
This formulation is not reflecting the interference pattern found when
considering electrons in the double slit experiment. We remark that $%
p_{AB}(x,t)$ should remind us of the probability we use to denote the
probability of the union of two mutually exclusive events $A$ and $B$:%
\begin{equation}
P(A\cup B)=P(A)+P(B).
\end{equation}

It is not possible to capture interference by superposing probability
distributions. Instead one could superpose probability waves, which we
denote as $\psi _{A}(x,t)$ for when slit $A$ is open and $\psi _{B}(x,t)$
for when slit $B$ is open. An alternative name for probability wave, which
is often used, is probability amplitude. The relationship between the
probability wave and the probability density is as follows: 
\begin{equation}
\left\vert \psi (x,t)\right\vert ^{2}\varpropto \text{probability density
function.}
\end{equation}

We remark that $\left\vert \psi (x,t)\right\vert ^{2}$ is obtained in
multiplying the probability wave (probability amplitude) with its complex
conjugate. Hence, we can write that: $\left\vert \psi (x,t)\right\vert
^{2}=\psi ^{\ast }\psi $. Please see below (just under (7)) where we provide
for an example.

Hence, we have that:

\begin{equation}
p_{A}(x,t)\varpropto \left\vert \psi _{A}(x,t)\right\vert ^{2}.
\end{equation}

Similarly for $p_{B}(x,t)$. Using (4) we can define the so called
\textquotedblleft quantum state\textquotedblright\ as the square root of the
probability density function. The quantum state, as per Khrennikov (2002),
\textquotedblleft is a purely mathematical quantity used to describe a
rather special behavior of probability densities of ensembles of systems
that are very sensitive to perturbations produced by
interactions.\textquotedblright

Since we are using waves, we can use superpositions. We can define: 
\begin{equation}
\psi _{AB}(x,t)=\psi _{A}(x,t)+\psi _{B}(x,t),
\end{equation}

where $\psi _{AB}(x,t)$ is the superposed state.

The above proportionality relationships can now be used to obtain:

\begin{equation}
p_{AB}(x,t)\varpropto \left\vert \psi _{A}(x,t)+\psi _{B}(x,t)\right\vert
^{2}.
\end{equation}

Before we move on, we need to introduce a basic notion from the area of
complex numbers. We know that a complex number $z$ can be denoted as $z=x+iy$%
, where $x$ is the real part and $y$ is the imaginary part. This same number 
$z$ can also be written as $z=re^{i\theta }$ where $r=\sqrt{x^{2}+y^{2}}$
and often $r$ can be denoted as $\left\vert z\right\vert $. The angle $%
\theta =\tan ^{-1}\frac{y}{x}$ and $x=r\cos \theta $ and $y=r\sin \theta $.
We also say that $\left\vert z\right\vert $ is the amplitude and $\theta $
is the phase.

We can write the wave function, which is a complex number, in exactly the
same way. Thus, the two components of a wave function are its phase and its
amplitude. The wave function can then be written as follows: 
\begin{equation}
\psi _{A}(x,t)=\left\vert \psi _{A}(x,t)\right\vert e^{iS_{A}(x,t)},
\end{equation}

where $S_{A}(x,t)$ is the phase of the wave function and $\left\vert \psi
_{A}(x,t)\right\vert $ is its amplitude. We write out $\psi _{B}(x,t)$ in
the same way.

We note that the complex conjugate, $\psi _{A}^{\ast }(x,t)$ is in this case
defined as: $\psi _{A}^{\ast }(x,t)=\left\vert \psi _{A}(x,t)\right\vert
e^{-iS_{A}(x,t)}.$

Substituting (7), and similarly for $\psi _{B}(x,t)$, in (6), we obtain:

\begin{equation}
p_{AB}(x,t)=\left\vert \psi _{A}(x,t)\right\vert ^{2}+\left\vert \psi
_{B}(x,t)\right\vert ^{2}+2\left\vert \psi _{A}(x,t)\right\vert \left\vert
\psi _{B}(x,t)\right\vert \cos (S_{A}-S_{B}).
\end{equation}

This is precisely the result which includes probability interference. I.e.
the additional term $2\left\vert \psi _{1}(x,t)\right\vert \left\vert \psi
_{1}(x,t)\right\vert \cos (S_{1}-S_{2})$ when not zero makes probability in
a quantum context, to be either sub-or super additive. For a much more in-
depth treatment on the comparison of (8) with (2), please see Khrennikov
(2007).

Khrennikov (2002) has indicated that quantum theory is a \textquotedblleft
special theory of statistical averages\textquotedblright .

\section{The sure-thing principle and the Ellsberg and Allais paradoxes}

The sure-thing principle is a key principle in utility theory. Utility
theory has as its primitive the so called utility function. The fundamental
question any economic theory textbook asks is whether, if good $x$ is
preferred over $y$, one can find an `if and only if' relationship with a
utility function such that the utility level of $x$ expressed through that
utility function is larger than the utility level of $y$ expressed through
that same utility function. The concept of \textit{expected} utility makes
up the `workhorse' of economic theory and its many applications. Expected
utility can be seen as the average value (probabilistically weighted) of
utility one gets through for instance playing a gamble. This concept will
become clear once we show how the Ellsberg and Allais' paradoxes can be
defined.

It needs to be stressed that economics has a variety of expected utility
theories. The most used (but the least realistic) expected utility approach
is the so called von Neumann-Morgenstern (1947) expected utility. The type
of probability used in formulating expected utility, in this model, can be
seen, as Kreps (1988) clearly indicates, as \textquotedblleft an objective -
externally imposed probability\textquotedblright . At the opposite end of
the expected utility theory spectrum, we can find the so called Savage
expected utility. This form of expected utility was developed by the famous
statistician Leonard Savage (1954). Probability in the Savage model is
entirely determined by the economic agent. Hence, in this model subjective
probability is used in the calculation of expected utility. A mixture of
objective and subjective probability is used in the Anscombe-Aumann (1963)
model. Aumann was the recipient of the recent Nobel Prize in economics.

The sure-thing principle is a key axiom in the Savage expected utility
approach. We can explain the principle in the following way. Let us imagine
we have experiment participants who are instructed to express a preference
over gambles $A$ and $B$ (set 1) and gambles $C$ and $D$ (set 2). First,
they express their choice of gamble $A$ over gamble $B.$ Second, they
express their choice of gamble $C$ over gamble $D$. On what basis do the
experiment participants express their choice? The participants are informed
of the payoff of the gambles in say, three states of nature; $s_{1},s_{2}$
and $s_{3}$. Assume that the payoff of gambles in set 1 for state $s_{3}$
are identical. Similarly, the payoff of gambles in set 2 for state $s_{3}$
are also identical. We note that the identical payoffs in $s_{3}$ for set 2
maybe different from the identical payoffs in $s_{3}$ for set 1. The
sure-thing principle says that the preference of the experiment participants
over the two gambles $A$ and $B$ (set 1) and $C$ and $D$ (set 2) will be
unaffected by the identical outcomes in state $s_{3}$. Hence, if we were to
swap the identical outcomes in $s_{3}$ for the gambles in set 2 with the
identical outcomes in $s_{3}$ for the gambles in set 1, this would NOT
affect the preference of the experiment participants. As is mentioned in
Busemeyer et al. (2006 and 2007), Shafir and Tversky (1992) have found in
many instances that players do violate this principle. In this paper we do
not want to detail the implications this violation has had on basic economic
theory models. An excellent source which treats this nevertheless important
issue is again Kreps (1988). Economic theory did find a response to this
defect. The core papers are by Gilbao and Schmeidler (1989) and Ghirardato
et al. (2004).

Now that we have illustrated the sure-thing principle, we can consider a
famous paradox, known also as the Ellsberg paradox, which implies a
violation of the sure-thing principle. Virtually any textbook in economic
theory will give an outline of this paradox. We describe it in the typical
way. Consider the following experiment. We have an urn with 30 red balls and
60 other balls (blue and green). We do not know the exact proportion of
green and blue balls. We consider four gambles and we ask experiment
participants to express a preference between gambles 1 and 2 and between
gambles 3 and 4. The gamble's payoffs are as follows.

\begin{enumerate}
\item Gamble 1 $(G1)$: you receive 1 unit of currency (uoc)\ if you draw a
red ball

\item Gamble 2 $(G2)$: you receive 1 unit of currency (uoc) if you draw a
blue ball

\item Gamble 3 $(G3)$: you receive 1 unit of currency (uoc) if you draw a
red or green ball

\item Gamble 4 $(G4)$: you receive 1 unit of currency (uoc) if you draw a
blue or green ball
\end{enumerate}

Most of the experiment participants (and this result has occurred in
repeated experiments) will prefer $G1$ over $G2$, $G1\succ G2$. The
intuition for such preference can be explained by the fact that one knows
the odds of winning in $G1$ (i.e. 1/3 probability) but in $G2$ one is unsure
about the odds. Participants in this experiment also indicated that $G4\succ
G3$. Here again, one knows the odds of winning in $G4$ are 2/3. However, one
is unsure about the odds in $G3$. Hence, the odds are ambiguous in $G2$ and $%
G3$.

This paradox clearly violates the sure-thing principle. This can be easily
shown if we use the following table to summarize the payoffs (in units of
currency):

\begin{center}
$%
\begin{array}{cccc}
& \text{Red} & \text{Blue} & \text{Green} \\ 
G1 & 1 & 0 & 0 \\ 
G2 & 0 & 1 & 0 \\ 
G3 & 1 & 0 & 1 \\ 
G4 & 0 & 1 & 1%
\end{array}%
$
\end{center}

The constant payoffs in set 1 ($G1$ and $G2$) and in set 2 $(G3$ and $G4$)
should not influence the preferences. Hence, if $G1\succ G2$ then, using the
sure-thing principle, it should be $G3\succ G4$. As indicated already above,
experiment participant will very often indicate $G3\prec G4$.

Let us denote the probability of drawing a red ball, as $p_{r}$. We know $%
p_{r}=1/3$. We can denote the other probabilities in a likewise fashion.
From the choice the experiment participants expressed one can see very
quickly that $G1\succ G2$ implies that $p_{r}>p_{b}$, if we assume that if
the units of currency one can gain would precisely coincide with the level
of utility (or satisfaction) of this gain. Similarly, the choice $G4\succ G3$
would now imply the opposite: $p_{b}>p_{r}$.

We round off this section with another paradox. This paradox is also widely
known in the economics literature and was first initiated by Economics Nobel
prize winner Maurice Allais. We follow here the example in Wolfram Mathworld
(2007). There are two experiments, consisting each of gambles $A$ and $B$
for experiment 1 and gambles $C$ and $D$ for experiment 2. We have again
four gambles. The payoff table, where we assume all numbers are in units of
currency, can be expressed as follows:

\begin{center}
$%
\begin{array}{ccccc}
gambles & [1,33] & 34 & [35,100] &  \\ 
A & 2500 & 0 & 2400 &  \\ 
B & 2400 & 2400 & 2400 &  \\ 
C & 2500 & 0 & 0 &  \\ 
D & 2400 & 2400 & 0 & 
\end{array}%
$
\end{center}

Drawing, in gamble $A$, balls with numbers in $[1,33]$ yields 2500$uoc$. If
the experiment participant were to draw ball number 34 (s)he would have a
payoff of 0$uoc.$ Finally, if the participant were to draw a ball number in $%
[35,100]$ it will yield 2400$uoc$. Gambles $A$ and $C$, without the $%
[35,100] $ event are identical. So are gambles $B$ and $D$ without the $%
[35,100]$ event. Thus, in this case, if test participants were to prefer $%
B\succ A$ they should also prefer $D\succ C.$ However, with the adding of
the $[35,100] $ event, individuals have shown in repeated experiments, to
exhibit the preference: $C\succ D$ and $B\succ A.$ If the experiment
participant were to have picked ball number `36' in both gambles $A$ and $B$
then by virtue of the identical outcome $(2400uoc$), we would have
indifference between gambles $A$ and $B.$ Similarly, if the experiment
participant, were to have picked ball `36' in both gambles $C$ and $D$, then
again there should be indifference between both gambles. Hence, the $%
[35,100] $ event is an independent event. It is intuitive to assume that if
we add an independent event to an identical experiment, the choice behavior
should not be influenced. Here, the expressed choice exhibited by the test
participants clearly contradicts this.

We can consider an interesting analogy between this paradox and the double
slit experiment. Pietro La Mura (2006) indicates that the particle in the
double slit experiment behaves like a decision maker violating the above
experiment. Says La Mura "why should it matter to an individual particle
which happens to go through the left slit, when determining where to scatter
later on, with what probability it could have gone through the right slit
instead?" Indeed the event of the particle going through the left slit
should be independent from the event of the particle to go through the right
slit. But that is of course not the case when there is probability
interference. Thus, if $B$ is preferred over $A$ then $D$ should be
preferred over $C$ and the action of the independent event should have no
influence. Thus, the probability of going through the right slit should be
of no importance to a particle going through the left slit. This is exactly
the influence of the independent event.

\section{Explaining the sure-thing principle with probability interference:
the Busemeyer et al. (2006, 2007) approach}

In their important papers Busemeyer et al. (2006 and 2007) discuss how the
violation of the sure-thing principle could be explained with the concept of
probability interference. The experiment they refer to consists of two
identical gambles. The gamble has two states of nature. One state of nature
corresponds to a winning amount of money and the other state to a losing
amount of money. Participants are informed they can play the game twice, in
sequence. In this experiment three situations are distinguished: a) the
participant is told (s)he won in the first play of the gamble; b) the
participant is told (s)he lost in the first play of the gamble; c) the
participant is told nothing as to the outcome of the first play of the
gamble. As indicated in Busemeyer et al. (2006 and 2007), if participants
prefer to play the second time the gamble, knowing that they won at the
first play, and if they prefer to play the second time the gamble knowing
they lost at the first play, then they should prefer to play a second time
even if they are not told whether they either lost or won at the first play.
This very intuitive prediction is entirely based on the sure-thing
principle. The study by Shafir and Tversky (1992), as quoted in the
Busemeyer et al. (2006 and 2007) paper, showed that 69\% of experiment
participants would go for a second play knowing that they won in the first
play. Furthermore, 59\% opted for a second play when they knew they had lost
in the first play. However, 36\% of the participants would NOT play for a
second time if they did not know the outcome of the first play of the
gamble. This behavior violates the sure-thing principle.

Busemeyer et al. (2006 and 2007) define two states of belief about the first
play that experiment participants can have: win or lose. There are two
states of action experiment participants can take: gamble or not. An
experiment participant can simultaneously have beliefs and actions. Those
produce four possible states, which are denoted in the typical quantum
physical `ket' way: $\left\{ |WG>,|WN>,|LG>,|LN>\right\} $. Each of the four
states, indicates the simultaneous belief experiment participants have in
for instance `winning (in the first play) of the gamble and having an
intention to gamble (in a second play)' $(|WG>)$ or in for instance `losing
(in the first play) but having no intention to gamble (in a second play)' $%
(|LN>)$. We mentioned in section two that in quantum physics we define a
probability wave or probability amplitude. The state function vector, in the
experimental setting here, is defined as: $\overrightarrow{\psi }=[\psi
_{_{WG}},\psi _{_{WN}},\psi _{_{LG}},\psi _{_{LN}}]$, where each $\psi _{..}$
is a probability amplitude. The $\overrightarrow{\psi }$ is of unit length.
We recall also from section two, that we defined $\left\vert \psi
\right\vert ^{2}$ (see (3)) as proportional to a probability density
function. We denote the probability value as: $\left\Vert \psi
_{..}\right\Vert ^{2}$. As an example $\left\Vert \psi _{_{WG}}\right\Vert
^{2}$ indicates the probability value of observing the state "winning and
gambling in a second play".

As is indicated in Busemeyer et al. (2006 and 2007), one can represent
`thought', which alters the state of the cognitive system, by a unitary
operator, $U$, such that $\overrightarrow{\varphi }=U.\overrightarrow{\psi }$%
. As an example (see Busemeyer et al. (2006 and 2007) for more examples) ,
if the experiment participant receives information she has lost in the first
gamble, then the operator $U$ is applied so as to create a new state price
vector: $\overrightarrow{\psi _{L}}=[0,0,\alpha _{_{L}},\beta _{_{L}}].$
Note that in the unknown case we have the superposition: $\psi _{_{U}}=\sqrt{%
p}\psi _{_{W}}+\sqrt{q}\psi _{_{L}}$, where $\sqrt{p}$ and $\sqrt{q}$ are
probability amplitudes.

Busemeyer et al. (2006 and 2007) now introduce the way players select a
strategy. As indicated above, a new state is formed through an operator $%
U_{t}$ for some period of time $t$, such that: $\overrightarrow{\varphi }%
=U_{t}.\overrightarrow{\psi }=\left[ \varphi _{_{WG}},\varphi
_{_{WN}},\varphi _{_{LG}},\varphi _{_{LN}}\right] $. The so called final
response probabilities are elements of $\overrightarrow{\phi }=M.%
\overrightarrow{\varphi }$, where $M$ is another operator. The elements of
this vector are the probabilities of playing a second time.

Hence, we can, as in Busemeyer et al. (2006 and 2007), define the density
function from which we can determine the total probability of gambling in
the second play. Using the concept of complex conjugate which we discussed
in section two (just under equation (7)), we obtain: $\left\vert \phi
\right\vert ^{2}=\phi ^{\ast }\phi $, where $\phi ^{\ast }$ is the complex
conjugate. The total probability of gambling the second time is: $\left\vert
\phi \right\vert ^{2}=\left( M\varphi \right) ^{\ast }(M\varphi )=\left\vert
M\varphi _{_{WG}}\right\vert ^{2}+\left\vert M\varphi _{_{LG}}\right\vert
^{2}$. Similarly, the total probability of not gambling the second time is: $%
\left\vert M\varphi _{_{WN}}\right\vert ^{2}+\left\vert M\varphi
_{_{LN}}\right\vert ^{2}$.

How can the probability interference term explain the sure-thing principle
violation? Using the development set out in this section, the probability of
winning is: $\phi _{W}^{\ast }\phi _{W}=\left( MU_{t}\psi _{_{W}}\right)
^{\ast }(MU_{t}\psi _{W})$ and similarly for losing: $\phi _{L}^{\ast }\phi
_{_{L}}=\left( MU_{t}\psi _{_{L}}\right) ^{\ast }(MU_{t}\psi _{_{L}})$.
Similarly Busemeyer et al. (2006 and 2007) show they can formulate the
probability in the case of the unknown condition, using the superposition $%
\psi _{_{U}}=\sqrt{p}\psi _{_{W}}+\sqrt{q}\psi _{_{L}}$. They get: $\phi
_{U}^{\ast }\phi _{U}=\left( MU_{t}\psi _{_{U}}\right) ^{\ast }(MU_{t}\psi
_{U})=$

\begin{equation}
\left( MU_{t}(\sqrt{p}\psi _{_{W}}+\sqrt{q}\psi _{_{L}})\right) ^{\ast
}(MU_{t}(\sqrt{p}\psi _{_{W}}+\sqrt{q}\psi _{_{L}}))
\end{equation}

and this product can be set equal to:

\begin{equation}
\left( \sqrt{p}\phi _{_{W}}+\sqrt{q}\phi _{_{L}}\right) ^{\ast }\left( \sqrt{%
p}\phi _{_{W}}+\sqrt{q}\phi _{_{L}}\right)
\end{equation}

and this is then equal to:

\begin{equation}
\left( p\phi _{_{W}}^{\ast }\phi _{_{W}}+q\phi _{_{L}}^{\ast }\phi
_{L}\right) +\sqrt{p}\sqrt{q}\phi _{_{W}}^{\ast }\phi _{_{L}}.
\end{equation}

What is key is now to remark that if one were to only consider:

\begin{equation}
\phi _{U}^{\ast }\phi _{U}=\left( p\phi _{_{W}}^{\ast }\phi _{_{W}}+q\phi
_{_{L}}^{\ast }\phi _{L}\right)
\end{equation}

then (12) would indicate that the probability of gambling for a second time,
when the experiment participant has no idea of whether he lost or won in the
first play, should be the average of the probabilities of gambling a second
time when the experiment participant knew that he respectively won and lost
in the first play. The experiments by Tversky and Shafir (1992) have clearly
shown that 36\% of the participant would not play for a second time when
they did not know whether they had lost or won in the first play. This falls
well below the 69\% of participants playing, knowing they had won in the
first play and the 59\% of the participants knowing they had lost in the
first play. Hence, it is clearly the interference term $\sqrt{p}\sqrt{q}\phi
_{_{W}}^{\ast }\phi _{_{L}}$, which was already covered in section 2 under
(8), which can explain the experimental observation of Tversky and Shafir
(1992).

\section{The concept of `ambiguity' in the Ellsberg paradox}

The Ellsberg paradox, as we have seen in section three of this paper
violates the sure-thing principle and it thereby contradicts expected
utility theory. Let us consider, like in Bossaerts at al. (2007), that the
red, green and blue balls are financial/economic securities. Examples of
securities are bonds, shares etc...Each security pays the same fixed amount
(1 unit of currency) according to the draw of the color. Hence, a `red'
security will be a security which pays 1 unit of currency if a red ball is
drawn from the urn of balls. The `red' security is risky in the sense that
the distribution of payoffs is known. However, the number of `green' and
`blue' securities are unknown in number. We say, as in Bossaerts et al.
(2007), that the `blue' and `green' securities are ambiguous. In other
words, the distribution of their payoffs, in the Ellsberg paradox is
unknown. We also reported the fact, in section three, that the probability
of drawing red is larger than the probability of drawing blue, $p_{r}>p_{b}$%
, when experiment participants expressed the preference: $G1\succ G2$.
However, when experiment participants expressed the preference of $G4\succ
G3 $, the opposite occurred: $p_{r}<p_{b}.$ If the price of the `red'
security, $q_{r}$, were to be tied to the knowledge of probability (and of
course prize winning or not), then clearly the price of the `red' security
should exceed the price of the `blue' security (in gambles 1 and 2): $%
q_{r}>q_{b}$. However, in gambles 3 and 4, the opposite effect occurs: $%
q_{r}<q_{b}$.

Let us now assume, still as in Bossaerts et al. (2007), that gambles 1 and 2
would refer to one market. Let us denote this market, as market $X$.
Moreover, let us assume that gambles 3 and 4 refer to another market, say
market $Y.$ It is quite straightforward to observe that one can
simultaneously buy a red security in market $Y$ and sell an identically same
red security in market $X$. Similarly, we can simultaneously buy the blue
security in the $X$ market and sell an identically same blue security in the 
$Y$ market. Such simultaneous buying and selling, without incurring risk, in
two different markets is also known as `arbitrage'.

Hence, the Ellsberg paradox is equivalent to the existence of an arbitrage
opportunity. As we have attempted to illustrate above, this paradox can be
explained via the probability interference argument. Hence, it may seem
reasonable to explain arbitrage via the probability interference argument.

Before entering into the next section, we need to stress that the concept of
`arbitrage' is an essential concept in the theory of security (or also
asset) pricing. For instance, the theory of derivative pricing (Black and
Scholes (1973)) is heavily dependent on the arbitrage concept. Most of the
theoretical work in asset pricing relies heavily on the so called
`non-arbitrage' theorem (Harrison and Kreps (1979)). This theorem, indicates
that for the case of a discrete state space, there will be no-arbitrage if
and only if there exists a set of risk-neutral probabilities. Risk-neutral
probabilities can be thought of as probabilities which allow the discounting
of a risky asset with the help of a risk free rate of interest. This
property is of capital importance in asset pricing. See Etheridge (2002) for
an excellent treatment. Kabanov and Stricker\ (2005) deal with the theorem
in the context of a continuous state space. Duffie (1996) discusses the
proof of the theorem. We propose that those risk-neutral probabilities could
be mapped onto the probabilities defined from $\left\Vert \psi \right\Vert
^{2}$, where the latter quantity was already discussed in section four
above. In fact, we can be even more explicit, by writing:

\begin{equation}
\left\Vert \psi (x)\right\Vert ^{2}=\int_{a_{1}}^{a_{2}}\left\vert \psi
(x)\right\vert ^{2}dx,
\end{equation}

where we have assumed time independence. For a set of bounds on the
integral, $\{a_{1},a_{2}\}$ we have a precise probability value on the left
hand side of the above equality. The probability density function is $%
\left\vert \psi (x)\right\vert ^{2}$. We now imagine that the set of
risk-neutral probabilities needed to ensure no arbitrage can be drawn, for a
given $\left\vert \psi (x)\right\vert ^{2}$, from the different sets of
bounds $\left\{ a_{i},a_{j}\right\} $ used in (13).

\section{Modelling information via the use of the wave function}

In this section we briefly exhibit how the wave function we defined in
section two, equation (7), can be used to model information. We recall that
the wave function was defined as: $\psi (x,t)=\left\vert \psi
(x,t)\right\vert e^{iS(x,t)}$, where $\left\vert \psi (x,t)\right\vert $ is
the amplitude of the wave function and $S(x,t)$ is the phase of the wave
function. We use $x$ and $t$ to denote respectively position and time.

The rationale for using such a wave function in the context of information,
is to convince the reader that a basic quantum mechanical tool like a wave
function can be used to model `ambiguity' (which we find back in the
Ellsberg paradox (please see section five above)) and `ignorance about
ambiguity'. We will discuss the relationship of the wave function with
ambiguity and `ignorance about ambiguity' in respectively sections seven and
eight below.

The use of a wave function as an information function was already proposed
in 1993 through the important work of Bohm and Hiley (1993). The authors
considered the information wave function as a so called `pilot wave'
(wherefrom the name of pilot wave theory) steering a particle. The pilot
wave function interpretation has deep roots in quantum physics. Prince Louis
de Broglie came up, as early as 1929, with his interpretation of the wave
function by saying that such function (p. 16 in Holland (1993)): "..not only
does...determine the likely location of a particle it also influences the
location by exerting a force on the orbit." David Bohm in two seminal papers
(1952) took up the thread where de Broglie had left off. We do not want to
go into any detail on this theory but it is sufficient to say that Bohm and
Hiley (1993) compare the pilot wave to a radio wave which steers a ship on
automatic pilot. Pioneering work in using this information wave function in
a macro-scopic (economics) setting was performed by Andrei Khrennikov (1999;
2002; 2004) and also by Olga Choustova (2006; 2007). Haven (2005; 2007) has
attempted to use this set up in a financial options (and more general
pricing) context.

We can use the information wave idea in the basic notion of probability
value we defined in equation (13) above. A change in information could be
reflected by a change in the functional form of the (information) wave
function, $\psi (x,t)$. This change would have as consequence that the
probability value would change because of a change in information. Hence, we
can obtain a set of probabilities which trigger arbitrage by simply changing
the information wave function and keeping the same bounds $\{a_{i},a_{j}\}$.
We come back to this issue in the next section.

\section{The wave function and ambiguity}

Let us re-visit once more the Ellsberg paradox of sections three and five.
We recall the payoffs:

\begin{enumerate}
\item Gamble 1 $(G1)$: you receive 1 unit of currency (uoc)\ if you draw a
red ball

\item Gamble 2 $(G2)$: you receive 1 unit of currency (uoc) if you draw a
blue ball

\item Gamble 3 $(G3)$: you receive 1 unit of currency (uoc) if you draw a
red or green ball

\item Gamble 4 $(G4)$: you receive 1 unit of currency (uoc) if you draw a
blue or green ball
\end{enumerate}

We had 30 red balls and 60 blue and green balls. We do not know the
proportions of blue or green balls.

An experiment participant may reason that it is `fair' to assume there are
an equal amount of green and blue balls. We note that in using this `fair'
argument ambiguity has been completely ruled out\footnote{%
Using the 50/50 rule would indicate one is using the principle of
insufficient reason (Kreps, p. 146) - which says: "if (one) has no reason to
suspect that one outcome is more likely than another, then by reasons of
symmetry the outcomes are equally likely and hence equally likely
probabilities can be ascribed to them." This argument can be challenged. For
instance, one could look at the uncertainty concerning the true value $p$ of
picking up a green or blue ball as a prior (Kreps, p. 149).}. Hence, the
chance of drawing any of the balls, in any of the four gambles is 30/90. In
this case we would be indifferent between gambles 1 and 2, and between
gambles 3 and 4.

Therefore, if the price of the red security, $q_{r}$, is again tied to the
knowledge of probability then the price of the red security should now be
equal to the price of the blue security (in gambles 1 and 2): $q_{r}=q_{b}$.
Furthermore, the price of the red security is equal to the price of the blue
security in gambles 3 and 4: $q_{r}=q_{b}$. Thus, with the 50/50 rule there
is no possibility for arbitrage: i.e. we can not simultaneously buy and sell
in two different markets and make a riskless profit. In fact, it is
precisely the existence of ambiguity which has created the arbitrage
opportunity.

Let us re-consider the role the wave function can play as an information
function in (13). We start from the assumption there is no arbitrage. In
this case the information wave function, $\psi (x,t)$, would contain a
precise piece of information. I.e. the precise piece of information would
be: \textquotedblleft we have a 50/50 chance of getting a blue or green ball
(i.e. security)\textquotedblright\ or alternatively \textquotedblleft the
price of the red security is equal to the price of the blue security in all
gambles\textquotedblright . Any departure from this, i.e. a change in the
information wave function, $\psi (x,t)$, which brings us back into
ambiguity, will yield arbitrage. Hence, a departure of the information wave
function from the 50/50 rule (or any other idiosyncratic - subjective
probability assessment), steers us away from expected utility and brings us
into ambiguity and therefore arbitrage.

\section{The wave function and ignorance about ambiguity}

We can go one step further by attempting to model ignorance about ambiguity.
To achieve this purpose we need to first inquire how ignorance could be
modelled with basic quantum mechanical tools. In a very interesting paper,
Franco (2007) provides for convincing arguments on how to model so called
`rational ignorance'. Rational ignorance comes into existence when the cost
of having to inform oneself about an issue outweighs the benefits of knowing
about that issue. Franco (2007) defines an opinion state, which is a qubit.
In a quantum state, there exists a linear superposition of 0 and 1:

\begin{equation}
a|0>+b|1>,
\end{equation}%
where $|0>=\left( 
\begin{array}{c}
1 \\ 
0%
\end{array}%
\right) $ and $|1>=\left( 
\begin{array}{c}
0 \\ 
1%
\end{array}%
\right) $ and $a,$ $b\in \mathbb{C}$. The superposition reflects the
existence of ignorance.

A so called `ket', $|s>$ , is a row vector and the dual vector is a so
called `bra', $<s^{\prime }|$. The inner product is denoted as $<s^{\prime
}|s>$: the probability amplitude. Furthermore, as expected, $\left\vert
<s^{\prime }|s>\right\vert ^{2}$ is the probability to get state $s^{\prime
} $ as a result of the measure on state $s|>$.

Franco (2007) uses the following well known quantum mechanics terms:

\begin{itemize}
\item a Hermitian operator\footnote{%
Hermiticity is the key property all quantum mechanical operators must
possess so that they can represent an observable (or physically measurable
quantity). Consider an operator $\widehat{Q}$, \ then this operator is
Hermitian if for any state function, $\psi ,$ we have that: $\int_{\text{all
space}}\psi _{1}\left[ \widehat{Q}\psi _{2}\right] dv=\int_{\text{all space}}%
\left[ \widehat{Q}\psi _{1}\right] \psi _{2}dv.$ Please see Morrison (1990)
for an excellent overview.} which corresponds to an observable quantity

\item eigenvalues are the measurable values of the observable

\item eigenvectors $|a>$, correspond to the quantum state (example: $|0>$
and $|1>$)
\end{itemize}

As in Franco (2007), three matrices which are Hermitian operators, are
introduced:

\begin{equation}
\sigma _{x}=\left[ 
\begin{array}{cc}
0 & 1 \\ 
1 & 0%
\end{array}%
\right] ;\sigma _{y}=\left[ 
\begin{array}{cc}
0 & -i \\ 
i & 0%
\end{array}%
\right] ;\sigma _{z}=\left[ 
\begin{array}{cc}
1 & 0 \\ 
0 & -1%
\end{array}%
\right]
\end{equation}

Those matrices, also known under the name of `Pauli matrices' have as
property that $\sigma _{i}^{2}=I=\left[ 
\begin{array}{cc}
1 & 0 \\ 
0 & 1%
\end{array}%
\right] .$ Also, the determinants of each $\sigma _{i}=-1$ and the trace
(i.e. the sum of the diagonal elements) is in each $\sigma _{i}=0$.

The eigenvalues of those operators define three orthonormal basises. Each
eigenvector of one basis is an equal superposition of the eigenvector of any
of the other basises. Those basises, Franco (2007) reports, are called
mutually unbiased. The eigenvectors are then respectively $\left\{
|0>,|1>\right\} ;$ $\left\{ \frac{|0>+|1>}{\sqrt{2}},\frac{|0>-|1>}{\sqrt{2}}%
\right\} ;$ $\left\{ \frac{|0>+i|1>}{\sqrt{2}},\frac{|0>-i|1>}{\sqrt{2}}%
\right\} $, where $i$ is a complex number. The Pauli matrices have
eigenvalues of $\pm $ 1/2. The physical realization of a qubit is provided
for by a spin 1/2 particle (f.i. an electron).

The observable quantities can be associated to questions such that they give
the answers 0 (false) or 1 (true) (the eigenvalues)$.$ The vectors $|0>$ and 
$|1>$ are associated to the answers $0$ and $1$ (true or false). The vectors
, $|0>$ and $|1>$ are the truth values.

In order to force the Pauli matrices to have eigenvalues $\pm $ $1$, Franco
(2007) considers the projector operator$:\overset{\symbol{94}}{\text{ }P}%
_{z}=(I-\sigma _{z})/2$ which yields eigenvalues of 0 and 1. The associated
eigenvectors are thus $|0>$ and $|1>$. The observable $\overset{\symbol{94}}{%
P}_{z}$ is, in the context of rational ignorance, associated to a first
question asked to participants.

Franco (2007) then defines two other observables$\overset{\symbol{94}}{,%
\text{ }P}_{y}=(I-\sigma _{y})/2$ and $\overset{\symbol{94}}{P}%
_{x}=(I-\sigma _{x})/2$ which can be associated with a second question. By
virtue of the construction of $\overset{\symbol{94}}{P}_{z}$, $\overset{%
\symbol{94}}{P}_{y}$,$\overset{\symbol{94}}{\text{ }P}_{x}$(the mutually
unbiasedness), the answers to those observables (questions) are
statistically independent.

As an example, Franco (2007), starts with question $\overset{\symbol{94}}{P}%
_{z}$ which has either 0 or 1 as an answer and the associated eigenvectors
are $|0>$ and $|1>$. Now consider another question, $\overset{\symbol{94}}{P}%
_{x}$, which has as associated eigenvectors $\frac{|0>+|1>}{\sqrt{2}}$or $%
\frac{|0>-|1>}{\sqrt{2}}$ for the 0 and 1 eigenvalues. Here the eigenvectors
are clearly superposed. They express thus ignorance. The inner product
calculates the probability amplitude (and the square is the probability). In
this case we obtain thus a probability of 1/2 for obtaining either 0 or 1.

Now assume, again as in Franco (2007) that we have experiment participants
who are first being asked question $\overset{\symbol{94}}{P}_{z}.$ This
operator is having as eigenvectors $|0>$ and $|1>$. There is no
superposition. Hence, no ignorance. Now assume, there is another question
posed to the experiment participants: question 2, $\overset{\symbol{94}}{P}%
_{x}$. The eigenvectors are now $\frac{|0>+|1>}{\sqrt{2}}$or $\frac{|0>-|1>}{%
\sqrt{2}}$. The probability amplitude in this case, can for instance be
calculated by using the inner product of $|0>$\ $($eigenvector of question
1) and $\frac{|0>\pm |1>}{\sqrt{2}}$\ (eigenvector of question 2), yielding
1/$\sqrt{2}$. This yields a probability of 1/2.\textbf{\ }

As is very clearly indicated in Franco (2007) when the experiment
participant gives an answer to the second question, the eigenvector $|0>$\
will collapse to say $\frac{|0>+|1>}{\sqrt{2}}$\ for the answer `0'. The
experiment participant has expressed ignorance since there is superposition.
If question 1 is asked now again, then what is the probability amplitude?%
\textbf{\ }The eigenvector $\frac{|0>+|1>}{\sqrt{2}}$\ will now collapse to
say $|1>$\ for the answer '1' and the inner product will be formed out of $%
\frac{|0>+|1>}{\sqrt{2}}$\ and $|1>$. As is aptly remarked in Franco (2007),
one observes now that the same question may get a different answer, since
there is a 50\% probability to get answer `1'\textbf{. }Hence, the context
of asking has changed the answer to the question.

The development above can be applied to arbitrage. In their quest to make
arbitrage profits, banks employ entire research departments to uncover such
opportunities. Proprietary data sets are developed to uncover arbitrage
opportunities. Let us denote the cost of such datasets as the price of
information. Using an analogy with the rational ignorance concept, how do we
know that information prices can outweigh the benefit of an arbitrage
opportunity? This is not an obvious question.

To revert back to the notion of ambiguity and arbitrage we discussed when we
considered the Ellsberg paradox, we could claim that the very existence of
an ambiguity in the financial markets leads to an arbitrage opportunity. In
the context of the Ellsberg paradox, rational ignorance would consist of not
believing that a potential set of arbitrage opportunities can be
economically viable for it to be investigated.

However, here we would like to tackle `ignorance about ambiguity'. Let us
think of some financial asset whose price may have the potential for
arbitrage taking. We have two questions, asked in sequence, to the research
department of a bank. The questions are answered with either false (`0') or
true (`1'). The probability of this is also assessed. For instance, in the
first question we ask whether, the price of the red security is higher than
the price of the blue security, $q_{r}>q_{b}$ in market $X.$ The potential
answer in a superposition of 0 and 1 will influence the probability as we
have seen above. In question 2, we ask whether $q_{r}<q_{b}$ in market $Y$.
If on both questions, the research department of the bank, answers without
any superposition, then this would indicate there is no ignorance at all
about ambiguity. If the answers are correct (i.e. they recognize the correct
inequality for each market $X$ and $Y)$ then there will exist an arbitrage
opportunity.

What if the research department of the bank answers the questions with
superposition? Clearly, there would be ignorance about ambiguity and hence
the bank would not realize an arbitrage opportunity.

We could translate this into the following equations. Using an extension on
(13) (see also Choustova (2006)):

\begin{equation}
\left\Vert \Psi \right\Vert ^{2}=\int_{q_{1}}^{q_{2}}\left\vert \Psi
(q)\right\vert ^{2}dM(q),
\end{equation}

where $M(q)$ is a probability measure and where $\left\{ q_{1},q_{2}\right\} 
$ are information prices$.$ We imagine that the research department of a
bank is researching an arbitrage opportunity. Hence, the set of information
prices now changes from $\left\{ q_{1},q_{2}\right\} \rightarrow \left\{
q_{1}^{\prime },q_{2}^{\prime }\right\} $. Assume we ask the two questions,
we described above, to the research department. We claim that the ignorance
about ambiguity can be captured by $M(q)$. So if for question 1, we obtain
non-superposed answers, then:

\begin{equation}
1=\int_{q_{1}}^{q_{2}}\left\vert \Psi (q)\right\vert ^{2}dM(q)
\end{equation}

For question 2, assume we have no ignorance about ambiguity then:

\begin{equation}
1=\int_{q_{1}^{\prime }}^{q_{2}^{\prime }}\left\vert \Psi (q)\right\vert
^{2}dM^{\prime }(q)
\end{equation}

, where $M(q)\neq M^{\prime }(q).$ As we have indicated at the end of
section five and at the end of section six, the arbitrage probability is
influenced by the information wave function. The existence of ambiguity, in
the context of the Ellsberg paradox, vouches for an arbitrage opportunity.
So the arbitrage probability is then:

\begin{equation}
1/x=\int_{q_{1}^{\prime }}^{q_{2}^{\prime }}\left\vert \Psi ^{\prime
}(q)\right\vert ^{2}dM^{\prime }(q)\text{, }x\geq 1
\end{equation}

, where $\left\vert \Psi ^{\prime }(q)\right\vert ^{2}\neq \left\vert \Psi
(q)\right\vert ^{2}$.

Thus, if we have that $\int_{q_{1}}^{q_{2}}\left\vert \Psi (q)\right\vert
^{2}dM(q)\neq \int_{q_{1}^{\prime }}^{q_{2}^{\prime }}\left\vert \Psi
(q)\right\vert ^{2}dM^{\prime }(q)$ then there is ignorance about ambiguity
and we can assume there is no arbitrage opportunity.

Thus, in summary, ignorance about ambiguity is reflected by a probability
which is dependent on the superposition of the true and false statements,
expressing ignorance about ambiguity.

\section{Experimental set up: how to test for probability interference in a
macro-scopic setting?}

The work by Busemeyer et al. (2006 and 2007) indicates that probability
interference can explain the violations of the sure-thing principle. The
Ellsberg paradox was a showcase paradox indicating a violation of the
sure-thing principle. We have tried to show how the information wave
function can be modelled in the context of the Ellsberg paradox. Following
the work of Franco (2007) basic quantum physical principles were used to
model rational ignorance. We used this set up to define ignorance about
ambiguity.

All of the presented material so far has tried to implicitly highlight the
importance of using quantum physical principles in instances of violations
of the sure-thing principle. The most important question, hence becomes
whether we can prove, or at least test for the existence of probability
interference in a macro-scopic setting?

\subsection{Using probability interference in social science: general set-up}

We follow in this subsection the work by Khrennikov (2002). Assume there
exists two mutually exclusive features $A$ and $B$. Each feature has a dual
outcome, `0' or `1'. The outcome `0' in features $A$ and $B$ is denoted as
respectively $a_{1}$ and $b_{1}$. Similarly, the outcome `1' in features $A$
and $B$ can be denoted as respectively $a_{2}$ and $b_{2}$. There exists an
ensemble of experiment participants, $\Sigma $ who have the same mental
state. The ensemble probability is denoted as $p_{j}^{a}$ and is defined as:

\begin{equation}
p_{j}^{a}=\frac{\text{number of results }a_{j}}{\text{total number of
elements in }\Sigma }.
\end{equation}

Similarly for $p_{j}^{b}$. A new ensemble $\Sigma $ needs to be prepared to
perform the measurement $p_{j}^{b}$.

Ensembles, $\Sigma _{i}^{a}$ and $\Sigma _{i}^{b};i=1,2$ need then to be
prepared and they have states corresponding to the values of $A=a_{j}$ and $%
B=b_{j};j=1,2$. The following probability is then defined:

\begin{equation}
p_{ij}^{a|b}=\frac{\text{number of results }a_{j}\text{ for the ensemble }%
\Sigma _{i}^{b}}{\text{total number of elements in }\Sigma _{i}^{b}}.
\end{equation}

Likewise for $p_{ij}^{b|a}$. From classical probability theory, total
probability is defined as:

\begin{equation}
p_{j}^{a}=p_{1}^{b}p_{1j}^{a|b}+p_{2}^{b}p_{2j}^{a|b};j=1,2.
\end{equation}

A likewise definition can be made for $p_{j}^{b}$.

In the presence of probability interference, one obtains (see also section 2
in this paper):

\begin{equation}
p_{j}^{a}=p_{1}^{b}p_{1j}^{a|b}+p_{2}^{b}p_{2j}^{a|b}+2\sqrt{%
p_{1}^{b}p_{2}^{b}p_{1j}^{a|b}p_{2j}^{a|b}}\cos \theta _{j};j=1,2;
\end{equation}

where $\cos \theta _{j}$ is defined as:

\begin{equation}
\cos \theta _{j}=\frac{p_{j}^{a}-p_{1}^{b}p_{1j}^{a|b}+p_{2}^{b}p_{2j}^{a|b}%
}{2\sqrt{p_{1}^{b}p_{2}^{b}p_{1j}^{a|b}p_{2j}^{a|b}}}.
\end{equation}

As we have already remarked in section 2, with equation (8), if the $\cos
\theta _{j}$ in (23) is non zero, then this would be indicative of the
existence of quantum-like behavior of cognitive systems. A likewise
definition can be made for $p_{j}^{b}$.

\subsection{Description and discussion of the proposed experiment}

In this subsection of the paper we will propose the conjecture we want to
test. We also briefly describe the test we envisage. We note that the test
description which we will explain below does not contain the minute details
of the experimental set up. We only provide for a fairly general discussion
of how we could possibly test our conjecture.

In basic terms the experiment deals with having experiment participants
recognize a list of songs from a pre-determined list of songs. The tempo of
each of the songs is distorted by either lenghtening or shortening the
tempo. We use the definition of tempo as in Levitin and Cook (1996):
\textquotedblleft ...the amount of time it takes a given note or the average
number of beats occurring in a given interval of time, usually beats per
minute.\textquotedblright\ Kuhn (1974) defines, beat tempo, in a very
similar way as: \textquotedblleft ...the rate of speed of a
composition.\textquotedblright

The time of listening exposure the test participants are allowed so as to
recognize each of the songs varies. The time of exposure is lengthened when
the tempo is increased. However, the time of exposure is shortened when the
tempo of songs is decreased. The rationale for this time variation is
discussed below.

The experiment participants are divided into several groups. We first want
to discuss the `normed' group. Levitin and Cook (1996) in an experiment on
tempo recognition asked experiment participants to sing tunes of songs they
themselves had picked from a list of available songs. They examined how well
the participants could mimic the tempo of each of the chosen songs. In their
experimental set up they formed a group of experiment participants (250
students) who would fill in questionnaires which would allow, what Levitin
and Cook (1996) called `norming'. This questionnaire, in the words of
Levitin and Cook (1996), \textquotedblleft asked them (the test
participants) to indicate songs that they knew well and could hear playing
in their heads.\textquotedblright\ Experiment participants in the `normed'
group, in our study, will do exactly that: indicate songs they know well. It
will be important to choose songs which we can term as being `best known'.
However, we may want to make sure we use songs from different music
categories. In a study by Collier and Collier (1994) the authors examine
tempo differences in different types of music (like jazz and other types).
The authors find that the tempo differences are quite tied to the music
genre in question. Furthermore, Attneave and Olson (1971) indicate that the
size of a music interval is dependent on the level of frequency of the music.

In the Levitin and Cook (1996) study, the 250 students we mentioned above,
were all taken from the Psychology 101 course. Those students were given the
questionnaire and from their answers the experimenters chose 58 CD's. It is
important to stress that our experiment proposal does not really follow the
Levitin and Cook (1996) study closely, since we ask instead from
participants to recognize songs we randomly play from the list we selected
from the choice of songs the `normed' group puts forward. In our
experimental set up, the experiment participants have to recognize songs we
play for them when tempo and time of exposure is varied. In the Levitin and
Cook (1996) study participants are asked to sing or hum a song for a time
the participant selects him/herself and the obtained tempo is then compared
with the copyrighted song's tempo.

Besides the normed group, we have two other groups of experiment
participants. We note that none of the members of the `normed' group will
participate in the experiment itself. We have therefore two other groups.
Group 1 whose participants are subjected to song excerpts where tempo
decreases are randomly injected in each song. The time of listening exposure
for each song is short. Participants in group 2 will be subjected to song
excerpts where tempo increases are randomly injected in each song. The time
of the listening exposure for each song is longer than in each of the songs
used in group 1.

The rationale for the variation of the exposure time comes from a study by
Kuhn (1974)\footnote{%
Also reported in the Levitin and Cook (1996) paper} where it was found that
test participants were better able to detect tempo decreases rather than
tempo increases. Kuhn (1974) indicates that a study by Farnsworth (1969)
showed that the \textquotedblleft listener is most likely to change
affective terms with which he describes a piece of music whenever its tempo
is appreciably slowed or hastened.\textquotedblright \footnote{%
Kuhn's (1974) study also discusses rythm besides beat. We omit it here.}
Kuhn (1974) in his experiment set up, where he solely used professional
musicians, showed that the beat tempos which had been decreased were
identified faster (in a statistically significant way) than beat tempos
which had been increased. Kuhn (1974) indicates those results also had been
obtained, independently, in a study by Drake (1968). In more general terms
the work by Drake and Botte (1993) addressed a fairly similar problem. They
had subjects listen to two identical sequences (except for their tempo
difference). One sequence had a higher tempo than the other. The subjects
were asked to indicate which sequence was faster.

We note that the tempo experiments by Kuhn (1974) did not make use of music.
However, the Kuhn (1988) study does provide for this. In this study mention
is made of the work by Geringer and Madsen (1984) which actually finds, in
the context of orchestral music, that tempo increases were more easily
identified than tempo decreases.

The Kuhn (1988) study provides for a rich context in which one can
appreciate how tempo changes can be affected by extraneous variables such as
melody activity and audible steady beats. The study finds that when melody
activity and audible steady beat were kept constant, the experiment
participants (the participants were students at an elementary school with a
very strong music programme) could distinguish well between slow and fast
tempi. However, experiment results in this study showed that melody activity
clearly affected tempo perception but there was much less certainty as to
how tempo recognition is affected when audible beat was considered. It needs
to be stressed that, according to Kuhn (1988) there exists a sizable
literature which documents there is ambiguity in tempo perception. Kuhn
(1988) cites work by (amongst others) Madsen, Duke and Geringer (1986) and
Wang and Salzberg (1984). One way out, as suggested in Kuhn (1988), is then
maybe to only use musicians as experiment participants in groups 1 and 2
(besides the normed group participants). Says Kuhn (1988) \textquotedblleft
some researchers believe that only trained, sophisticated musicians can
correctly interpret and perceive `tempo' as it is usually
defined.\textquotedblright

Khrennikov and Haven (2006) have described an experiment where we can test
for the same type of variables (time and degree of deformation) by using
instead a database of pictures. This experiment formed part of a funding bid
the authors wrote for the Fund for Scientific Research (FWO)\footnote{%
`FWO' is the abbreviation in Dutch for "Fonds voor Wetenschappelijk
Onderzoek".} of the Flemish Government (Belgium)\footnote{%
We could also imagine an experiment where we test for deformations of both
verbal and photograpic memory.}.

Two features of the song recognition are being compared:

\begin{enumerate}
\item time of processing of the songs (see also section 9.4. for more
discussion on this parameter)

\item the ability to recognize a song, $S_{0}$ by analyzing a deformation $S$
of it (see also section 9.4.)
\end{enumerate}

We denote the time of processing of the song with the variable $t$. The
ability to recognize the song with tempo deformations which are either
increasing or decreasing, is denoted with the variable $a$.

The conjecture we want to test is:

\begin{conjecture}
$t$ and $a$ are complementary.
\end{conjecture}

\subsubsection{Preparation}

The state preparation of the experiment can be described as follows. The
experimental context (state) $C$ is given by a sequence of songs, $%
S_{1},...,S_{m}.$ The experiment participants form a group $G$ and each of
the participants are exposed to all the songs (from the distillated list of
songs taken from the normed group) excerpts they will hear. The context
allows thus the experiment participants to learn the songs (some of the
songs they surely will know, but some they may not know as well (or not at
all)).

In the Levitin and Cook (1996) study only 46 students were selected for the
tempo recognition experiment. After learning has taken place, the group $G$
is randomly divided into two equal subgroups $G_{1}$ and $G_{2}.$

\subsubsection{First experiment: song tempo decreases and short exposure time%
}

A first experiment is performed with experiment participants from group $%
G_{1}.$ As in Levitin and Cook (1996), test participants are seated in a so
called sound attenuation booth. Each of the participants is then subjected
to a battery of songs (selected from our `best' songs (from the normed
group)) played sequentially. Each of the songs is played for the same amount
of time. In this first experiment, the tempo of each of the songs is
decreased and the exposure time (to listen to the song) is kept uniformly
short\footnote{%
The `shortness' of the exposure time is obviously a calibration issue.
Please see below for some beginning discussion on this topic.}. Experiment
participants must choose from the full list of songs which song they are
listening to. Songs in the full list are indicated (alphabetically) by the
name of the singer or composer. Next to the authors' name is the title of
the song and the record company owning the copyright of the song. We also
note that experiment participants need to pick the song from the list within
a uniformly prescribed time interval. This is a possible issue of
contention. Kuhn (1974) reports that professional musicians need less time
to respond to a tempo change and they make also less mistakes than
non-professional musicians. A study related to this issue is Kuhn's 1975
paper which reports that on a longitudonal basis, test participants (of all
walks of life) when asked to keep a constant tempo, seem to instead increase
tempo. Since our study would not necessarily contain test participants who
are only musicians, setting the response time too short could thus create a
negative bias.

The low tempo songs $S_{1}^{\prime },...,S_{m}^{\prime }$ are deformations
of the original songs $S_{1},...,S_{m}$. The tempo change could be fine
tuned in the way Madsen and Graham (1970)(see also Kuhn (1974)) have
proposed. Those authors propose a modulation rate (of tempo) of one beat per
minute change every second.

Experiment participants in group $G_{1}$ are subjected to listening to $%
S_{1}^{\prime },...,S_{m}^{\prime }$ \textit{and }also to listening to a few
other songs which have not been part of the training sample. We denote this
set of songs as: $S_{G_{1}}.$ The width of the time window, $w$ is a
parameter of the experiment.

The task the experiment participants have to fulfill is to indicate whether
they either recognize or not the songs $S_{G_{1}}$ as modifications of the
songs $S_{1},...,S_{m}$. We can make this experiment a little more
sophisticated by using the procedure Levitt (1971) used. See also Drake and
Botte (1993) where the Levitt approach is described. Levitt (1971) would
decrease the tempo difference between two subsequent sequences by 1\%, if
the subjects gave two correct answers (to the two prior sequences). In some
sense the tempo decrease compensates for the shorter period which is given.
We could also make it increasingly harder for test participants by instead
increasing the tempo for every correct answer. We do not follow this
strategy in this paper.

Let $\omega $ be an experiment participant from group $G_{1}$ performing
this task. We set $t(\omega )=1$ if $\omega $ was able to give the correct
answers for $x\%$ of the songs in $S_{G_{1}}$. And $t(\omega )=0$ in the
opposite case.

We now find probabilities $P(t=1|C)$ and $P(t=0|C)$ through counting numbers
of experiment participants who gave answers $t=1$ and $t=0,$ respectively.
We denote respective subgroups of experiment participants by $G_{1}(t=1)$
and $G_{1}(t=0),$ respectively. The first subgroup consists of experiment
participants who have the ability to perform song recognition (with tempo
decrease) `quickly' and the second subgroup consists of experiment
participants who do not possess that feature.

\subsubsection{Second experiment: song tempo increase and long exposure time}

The second experiment is performed with experiment participants from the
group $G_{2}$ as well as the subgroups $G_{1}(t=0)$ and $G_{1}(t=1)$ of the
first subgroup $G_{1}.$

The song deformations now are based on song tempo increases which as per the
experiment of Kuhn (1974), are deformations which are harder to recognize.
We denote those deformations as $S_{1}^{\prime \prime },...,S_{m}^{\prime
\prime }$ of initial songs $S_{1},...,S_{m}$. As in the first experiment,
`unknown' songs will be added to the group of songs. We denote the set of
essentially deformed songs (with the `unknown' songs): $S_{G_{2}}$.

The task the experiment participants have to fulfill is identical to the
task described in experiment one: can the participants recognize the songs
in $S_{G_{2}}$ as modifications of the initial songs $S_{1},...,S_{m}$?

The time window in experiment 2 is now longer than in experiment 1. The
width of the time window, $w$ is a parameter of the experiment.

Let $\omega $ be an experiment participant performing this task. We set $%
a(\omega )=1$ if $\omega $ was able to give the correct answers for $x\%$ of
images in the series. And $a(\omega )=0$ in the opposite case.

We now find probabilities $P(a=1|C)$ and $P(a=0|C)$ through counting numbers
of experiment participants in the group $G_{2}$ who gave answers $a=1$ and $%
a=0,$ respectively. We denote respective subgroups of experiment
participants by $G_{2}(a=1)$ and $G_{2}(a=0),$ respectively. The first
subgroup consists of experiment participants who have the ability to perform
song recognition (with tempo increase) quickly and the second subgroup
consists of experiment participants who do not possess that feature. Because
of the tempo increase, we note that the emphasize in experiment two is on
the carefulness of recognizing the deformation of a song.

Here again, we could use the Levitt (1971) approach. We can increase (or
decrease) with 1\% the tempo between subsequent songs if the test
participant has given two correct answers.

\subsection{Sub (super) additivity}

Using the above experiment, we find probabilities $P(a=\beta |t=\alpha )$; $%
\alpha ,\beta =0,1,$ by counting the number of people in the group $%
G_{1}(t=\alpha )$ who gave the answer $a=\beta .$

After this we calculate the coefficient $\lambda $ (which is the third term
in equation (23)) and we find the angle $\theta $ which gives us the measure
of complementarity of variables $t$ and $a.$ If $\lambda >1$, we would find
experimental evidence of probabilistic behavior which is neither quantum nor
classical. We could also consider $\lambda (w)$ and even make $\lambda $
dependent on both $w$ and the degree of deformation. This degree of
deformation could be expressed by using the tempo measure defined by Levitin
and Cook (1996).

We note that sub-additive probability is a concept which has already been
studied in psychology. It has been shown that when experiment participants
have to express their degree of beliefs on a [0,1] interval, probabilistic
additivity will be violated in many cases and sub-additivity obtains. See
Bearden et al. (2005) for a good overview. Bearden et al. (2005) also
indicate that such sub-additivity has been obtained with experiment
participants belonging to various industry groups, such as option traders
for instance (Fox et al. (1996)). The key work pertaining to the issue of
sub-additivity in psychology is by Tversky and Koehler (1994) and
Rottenstreich and Tversky (1997). Their theory, also known under the name of
`Support Theory' is in the words of Tversky and Koehler (1994)
\textquotedblleft ...a theory in which the judged probability of an event
depends on the explicitness of its description.\textquotedblright\ In other
words, it is not the event which is important as such but its description.
In Tversky and Koehler (1994) the authors highlight the `current' state of
affairs (Anno 1994) on the various interpretations subjective probability
may have. Amongst the interpretations is Zadeh's possibility theory (1978)
and the upper and lower probability approach of Suppes (1974). The paper by
Dubois and Prade (1988), also mentioned in the Tversky and Koehler (1994)
article, provides for an excellent overview on non-additive probability
approaches.

\subsection{Complementarity of $t$ and $a$}

Can the conjecture 9.1. we posed at the beginning of this section make sense
from a quantum mechanical point of view? As is well known, there does not
exist a quantum operator on time at all! As we have remarked in Khrennikov
and Haven (2006) there exist a very well known test of complementarity,
based on the Heisenberg uncertainty relation which says that the product of
the uncertainties in the determination of position ($\Delta p=\sqrt{%
E(p-Ep)^{2}}$) and momentum ($\Delta p=\sqrt{E(p-Ep)^{2}}$) is weakly larger
than the Planck constant divided by two. However, as we remarked in
Khrennikov and Haven (2006), this method can not be applied for discrete
variables, since if operators have discrete spectra there exist eigenstates
(corresponding to eigenvalues) and their standard deviation equals to zero.
Hence, it is not possible to find an uncertainty principle of the type: $%
\Delta t.\Delta a=h_{\mathrm{cogn}}/2,$ where $h_{\mathrm{cogn}}$ \ is an
analog of the Planck constant. Discussions on analogs of Planck constants in
financial contexts and other non-quantum physical contexts were provided for
in Choustova (2006; 2007), Khrennikov (2004) and Haven (2005; 2007).

The very objective of our experiment is just to overcome those problems of
measurement and hence to be able to test for complementarity on the $t$ and $%
a$ variables.

\section{Conclusion}

In this paper we have tried to show how probability interference, which is
an essential quantum physical concept can be found back in a macro-scopic
setting, more precisely in a social science setting. Such interference can
help in explaining major economic paradoxes and the very paradoxes
themselves have quite some interesting ramifications. Therefore, testing for
the presence of such interference on a macro-scopic scale is an interesting
challenge. In this paper we provide for such a possible experiment.

\newpage

\section{Bibliography}

\begin{enumerate}
\item Anscombe F. , Aumann R. (1963), A definition of subjective
probability; \textit{Annals of Mathematical Statistics}; 34, 199-205.

\item Attneave F.; Olson R.K. (1971), Pitch as a medium: a new approach to
psychological scaling; 84, 147-166

\item Bearden J.N., Wallsten T., Fox C. (2005), Error and subadditivity: a
stochastic model of subadditivity, University of Arizona - Department of
Management and Policy.

\item Black F., Scholes M. (1973), The pricing of options and corporate
liabilities; \textit{Journal of Political Economy}; 81, 637-654.

\item Bohm D. and Hiley B. (1993), \textit{The Undivided Universe}, New
York: Routledge, 1-397.

\item Bohm D. (1987), \textit{Hidden Variables and the Implicate Order} in
Quantum Implications: Essay in Honour of D. Bohm, edited by B. Hiley and F.
Peat, New York: Routledge, 33-45.

\item Bohm, D. (1952). A suggested interpretation of the quantum theory in
terms of `hidden' variables, Part I and II. \textit{Physical Review} \textbf{%
85, }166-193.

\item Bossaerts P, P. Ghirardato, S. Guarnaschelli and W. Zame (2007);
Prices and Allocations in Asset Markets with Heterogeneous Attitudes Towards
Ambiguity; Working paper; \textit{California Institute of Technology}.

\item Busemeyer J. and Wang Z. (2007), Quantum information processing
explanation for interactions between inferences and decisions; Papers from
the AAAI Spring Symposium (Stanford University); 91-97

\item Busemeyer J. and Wang Z \ and Townsend J.T. (2006), Quantum dynamics
of human decision making, \textit{Journal of Mathematical Psychology} 50
(3): 220-241.

\item Choustova, O. (2006), Quantum Bohmian model for financial markets. 
\textit{Physica A} 374\textbf{, }304-314.

\item Choustova O. (2007), Toward quantum-like modelling of financial
processes\textit{; Journal of Physics: Conference Series }70 (38pp)

\item Collier G and Collier J.L. (1994), An exploration of the use of tempo
in jazz; \textit{Music Perception}; 11(3); 219-242

\item Conte E. , Todarello O., Federici A., Vitiello F., Lopane M.,
Khrennikov A. (2004), \textit{A preliminary evidence of quantum-like
behaviour in measurements of mental states}; Proc. Int. Conf. Quantum
Theory: Reconsideration of Foundations. Ser. Math. Modelling in Phys.,
Engin., and Cogn. Sc., vol. 10, 679-702, V\"{a}xj\"{o} Univ. Press.

\item Drake A. H. (1968), An experimental study of selected variables in the
performance of musical durational notation; \textit{Journal of Research in
Music Education}; vol. 16, 329-338

\item Drake C., Botte M.C. (1993), Tempo sensitivity in auditory sequences:
evidence for a multiple look model; \textit{Perception and Psychophysics};
54(3), 277-286

\item Dubois D., Prade H. (1988) , Modelling and inductive inference: a
survey of recent non-additive probability systems; \textit{Acta Psychologica}%
; 68, 53-78.

\item Duffie, D. (1996). \textit{Dynamic Asset Pricing Theory}, Princeton
University Press, Princeton.

\item Ellsberg, D. (1961), Risk, Ambiguity and the Savage Axioms;\ \textit{%
Quarterly Journal of Economics} 75, 643-669, 1961

\item Etheridge, A. (2002). \textit{A Course in Financial Calculus},
Cambridge University Press, Cambridge.

\item Farnsworth P.R. (1969). \textit{The Social Psychology of Music}, Iowa
State University Press, Iowa.

\item Fox C., Rogers B., Tversky A. (1996), Option traders exhibit
subadditive decision weights; \textit{Journal of Risk and Uncertainty; }13,
5-17.

\item Franco, R. (2007). Quantum mechanics and rational ignorance. \textit{%
arXiv:physics/0702163v1}

\item Geringer, J. and Madsen C.M. (1984). Pitch and tempo discrimination in
recorded orchestral music among musicians and nonmusicians; \textit{Journal
of Research in Music Education} 32, 195-204.

\item Ghirardato, P., Maccheroni F. and Marinacci M. (2004)
\textquotedblleft Differentiating Ambiguity and Ambiguity
Attitude.\textquotedblright\ \textit{Journal of Economic Theory} 118,
133-173.

\item Gilboa, I. and D. Schmeidler (1989) Maxmin Expected Utility with a
Non-Unique Prior; \textit{Journal of Mathematical Economics} 18, 141-153.

\item Harrison, J.M. and Kreps, D. M. (1979). Martingales and arbitrage in
multiperiod securities markets. \textit{Journal of Economic Theory }\textbf{%
20, }381-408.

\item Haven, E. (2005); Pilot-wave theory and financial option pricing; 
\textit{International Journal of Theoretical Physics} 44 (11), 1957-1962

\item Haven, E. (2007). Private information and the `information function':
a survey of possible uses. \textit{Theory and Decision}, forthcoming

\item Haven, E. (2007). The Variation of Financial Arbitrage via the Use of
an Information Wave Function; \textit{International Journal of Theoretical
Physics}, forthcoming

\item Holland, P. (1993). \textit{The quantum theory of motion: an account
of the de Broglie-Bohm causal interpretation of quantum mechanics.}
Cambridge University Press, Cambridge.

\item Kabanov, Yu. and Stricker, C. (2005). Remarks on the true no-arbitrage
property. \textit{S\'{e}minaire de Probabilit\'{e}s XXXVIII - }Lecture Notes
in Mathematics \textbf{1857}, 186-194, Springer.

\item Khrennikov, A. Yu. (1999). Classical and quantum mechanics on
information spaces with applications to cognitive, psychological, social and
anomalous phenomena. \textit{Foundations of Physics} \textbf{29}, 1065-1098.

\item Khrennikov, A. (2002) \textit{On the cognitive experiments to test
quantum-like behavior of mind}; Reports from V\"{a}xj\"{o} University -
Mathematics, Natural Sciences and Technology, Nr. 7.

\item Khrennikov, A. (2004) \textit{Information dynamics in cognitive,
psychological and anomalous phenomena}; Ser. Fundamental Theories of
Physics, v. 138, Kluwer, Dordrecht.

\item Khrennikov, A. (2007). Classical and quantum randomness and the
financial market. \textit{arXiv: 0704.2865v1 [math.PR]}

\item Khrennikov A., and E. Haven (2006) ; Does probability interference
exist in social science? \textit{Foundations of Probability and Physics -4
(G. Adenier, A. Khrennikov, C. Fuchs - Eds), AIP Conference Proceedings};
899; 299-309.

\item Kreps D. (1988), \textit{Notes on the theory of choice}, Westview
Press (Colorado-Boulder).

\item Kuhn T.L. (1974); Discrimination of modulated beat tempo by
professional musicians; \textit{Journal of Research in Music Education}; 22;
270-277.

\item Kuhn T.L. (1975); Effect of notational values, age, and example length
on tempo performance accuracy; \textit{Journal of Research in Music Education%
}; 23 (3); 203-210.

\item Kuhn T. L.; Booth, G. D. (1988); The effect of melodic activity, tempo
change, and audible beat on tempo perception of elementary school students; 
\textit{Journal of Research in Music Education}, 1988, 36, 140-155.

\item La Mura, P. (2006), Projective Expected Utility, \textit{Mimeo},
Leipzig Graduate School of Management.

\item Levitin D., Cook P.R. (1996), Memory for musical tempo: additional
evidence that auditory memory is absolute'; \textit{Perception and
Psychophysics}; 58; 927-935.

\item Levitt H. (1971), Transformed up-down methods in psychoacoustics; 
\textit{Journal of the Acoustical Society of America}; 49; 467-477

\item Madsen C.K., Duke R.A. and Geringer J.M. (1986) The effect of speed
alterations on tempo note selection; \textit{Journal of Research in Music
Education}, 34; 101-110

\item Madsen C. K. and Graham R. (1970) The effect of integrated schools on
performance of selected music tasks of black and white students; \textit{%
Music Educator's National Conference}; Chicago; Illinois

\item Morrison M. (1990), \textit{Understanding quantum physics; }%
Prentice-Hall.

\item Rottenstreich Y., Tversky A. (1997) Unpacking, repacking and
anchoring: advances in support theory; \textit{Psychological Review}; 104,
406-415.

\item Savage L.J. , \textit{The Foundations of Statistics}; NY, Wiley, 1954.

\item Shafir E and Tversky A. (1992); Thinking through uncertainty:
non-consequential reasoning and choice; \textit{Cognitive Psychology,}24,
449-474.

\item Suppes P. (1974), The measurement of belief; \textit{Journal of the
Royal Statistical Society B}; 36, 160-191.

\item Tversky A., Koehler D. (1994), Support theory: a nonexistential
representation of subjective probability; \textit{Psychological Review};
101, 547-567.

\item von Neumann J. , Morgenstern O. (1947), \textit{Theory of games and
economic behavior}; Princeton University Press.

\item Wang C.C. and Salzberg R.S. (1984), Discrimination of modulated music
tempo by strong students; \textit{Journal of Research in Music Education},
32, 123-131

\item Wolfram MathWorld; http://mathworld.wolfram.com/AllaisParadox.html

\item Zadeh L. (1978), Fuzzy sets as a basis for a theory of possibility; 
\textit{Fuzzy Sets and Systems}; 1, 3-28.
\end{enumerate}

\end{document}